# An Experimental and Theoretical Study of the Variation of 4f Hybridization Across the $La_{1-x}Ce_xIn_3$ Series

D. Gout,*,† O. Gourdon,† E. D. Bauer,‡ F. Ronning,‡ J. D. Thompson,‡ and Th. Proffen†

*Los Alamos Neutron Science Center and Materials Physics and Applications Division, Los Alamos National Laboratory, Los Alamos, New Mexico 87545*

Crystal structures of a series of $La_{1-x}Ce_xIn_3$ ($x = 0.02, 0.2, 0.5,$ or $0.8$) intermetallic compounds have been investigated by both neutron and X-ray diffraction, and their physical properties have been characterized by magnetic susceptibility and specific heat measurements. Our results emphasize atypical atomic displacement parameters (ADP) for the In and the rare-earth sites. Depending on the $x$ value, the In ADP presents either an "ellipsoidal" elongation (La-rich compounds) or a "butterfly-like" distortion (Ce-rich compounds). These deformations have been understood by theoretical techniques based on the band theory and are the result of hybridization between conduction electrons and 4f-electrons.

## Introduction

The rare earth metal (RE) $In_3$ intermetallic compounds (where RE = La or Ce) with the cubic $AuCu_3$ structure type have been the subject of many experimental and theoretical investigations. $CeIn_3$ is a member of heavy-electron materials that display fascinating interactions between the often antagonist phenomena of superconductivity and (anti)ferromagnetism. Indeed, $CeIn_3$ undergoes a transition from an antiferromagnetic (AFM) state at ambient pressure ($T_N = 10$ K) to a superconducting state ($T_c = 0.15$ K) at a critical pressure $P_c = 2.8$ GPa at which long-range magnetic order vanishes, suggesting that this material may exhibit magnetically mediated superconductivity.[1]

In addition to its intrinsic interest, $CeIn_3$ also is the structural building block of the tetragonal family of heavy-electron compounds $CeMIn_5$, where M = Co, Rh, and Ir. These materials can be viewed as alternating layers of weakly distorted $CeIn_3$ cuboctahedra and rectangular parallelepipeds ($MIn_2$) stacked sequentially along the $c$-axis (Figure 1).[2,3]

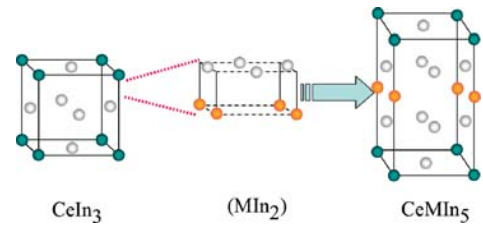

**Figure 1.** Schematic representation of the $CeMIn_5$ structure viewed as intergrowth of distorted $CeIn_3$ cuboctahedra and rectangular parallelepipeds ($MIn_2$).

Depending on the transition metal element M, these compounds are unconventional superconductors (M = Co or Ir)[4] or antiferromagnetic (M = Rh) at atmospheric pressure but become superconducting under applied pressure.[5] Detailed studies of these compounds suggest that their electronic ground-state is controlled by hybridization of cerium's 4f electron with ligand electrons provided by the M element[6] and leads to dominantly itinerant character of 4f electrons for M = Co or Ir and dominantly localized character when M = Rh.[7] How this hybridization evolves with temperature has been explored[8] by diluting the Ce atoms in $CeCoIn_5$ and CeIrIn$_5$ with La, which contains no 4f electrons. Analysis of these dilution studies suggest that, at temperatures below about 50 K, hybridization leads to two components in the f-electron degrees of freedom, one that is more band-like and one that is more localized-like, but presently, there is no theoretical model that accounts for these behaviors. Exploring the role of 4f hybridization in the parent compounds $La_{1-x}Ce_xIn_3$ will be important for guiding an interpretation of this two-component behavior in CeMIn$_5$.

**Table 1.** Compositions Targeted and Electron Microprobe Analyses for $La_{1-x}Ce_xIn_3$ ($x = 0.02, 0.2, 0.5,$ and $0.8$).

| targeted compositions | $La_{0.98}Ce_{0.02}In_3$ | $La_{0.8}Ce_{0.2}In_3$ | $La_{0.5}Ce_{0.5}In_3$ | $La_{0.2}Ce_{0.8}In_3$ |
| --- | --- | --- | --- | --- |
| electron microprobe analyses | $La_{0.97(1)}Ce_{0.03(1)}In_3$ | $La_{0.79(1)}Ce_{0.21(1)}In_3$ | $La_{0.51(1)}Ce_{0.49(1)}In_3$ | $La_{0.22(1)}Ce_{0.78(1)}In_3$ |

The electronic structure of RE In$_3$ compounds has been intensively studied by a number of researchers during the few last decades. Their work reasonably describes the role played by the f electrons in regards to the physical phenomena but mainly focuses on the topology of the Fermi surface of these alloys, especially how the surface is modified in LaIn$_3$ compared to CeIn$_3$ and how pressure can imply the superconductivity.

However, during these past years, the detailed crystal structure of these intermetallic compounds, and therefore the atomic arrangements at the origin of these properties, has been overlooked. Whereas the majority of related systems, such as CePd$_2$Si$_2$, CeRh$_2$Si$_2$, or UGe$_2$, possess a tetragonal or orthorhombic unit cell, with few nonequivalent atoms per primitive cell, RE In$_3$ crystallizes in the simple cubic AuCu$_3$ structure type and so presents an instructive case for studying these phenomena.

In this work, we present a reinvestigation of the local and spatial averaged crystal structure of a series of $La_{1-x}Ce_xIn_3$ ($x = 0.02, 0.2, 0.5,$ or $0.8$) intermetallics compounds using both neutron and X-ray powder diffraction. Local structure analyses have been investigated using pair distribution function (PDF) analyses, whereas the averaged crystal structure has been obtained by traditional Rievteld refinements. Our refinements emphasize atypical anisotropic displacement parameters (ADPs) for the In and the rare-earth (RE) sites, which lead to a certain critique of the cubic symmetry for these phases. The In ADP presents either an "ellipsoidal" elongation or a "butterfly-like" distortion, depending on the Ce/La concentration. Thereafter, tight-binding-linear muffin-tin orbital-atomic-spheres approximation (TB-LMTO-ASA) calculations were performed on $La_{1-x}Ce_xIn_3$ ordered models to interpret our structural refinements.

### Experimental Section

**Synthesis, Chemical Analysis, and Characterization.** The $La_{1-x}Ce_xIn_3$ samples with nominal composition between $x = 0$ and 1 were prepared by placing stoichiometric amounts of Ce (ingot, Ames Laboratory, 99.99%), La (ingot, Ames Laboratory, 99.99%), and In (ingot, Alfa Aesar, 99.999%) in a quartz tube sealed under vacuum. Particular care was taken to avoid using oxidized elements as this affects the stoichiometry. The materials were heated to 1050 °C and were kept there for 4 h, followed by slow cooling to 600 at 5 °C/h, at which point the excess In flux was removed using a centrifuge.

Electron microprobe analyses were performed using a JEOL JXA-8200 and with the pure elements as well as CeIn$_3$ as standards to obtain quantitative values. These analytical results are summarized in Table 1, which show excellent agreement with both the loaded compositions as well as those refined from subsequent diffraction experiments.

**Physical Characterization.** Magnetic susceptibility measurements were performed in a magnetic field of 1 kOe from 1.8 to 350 K using a Quantum Design MPMS magnetometer. Specific heat was measured in a Quantum Design PPMS system at temperatures from 0.4 to 20 K.

**Physical Properties. Magnetic Susceptibility.** The inverse magnetic susceptibility of $La_{1-x}Ce_xIn_3$ compounds is plotted in Figure 2a for samples with $x \geq 0.2$. In these data, the mole fraction contribution to the susceptibility from LaIn$_3$, which is a weakly temperature-dependent Pauli paramagnet with $\chi \approx 4 \times 10^{-4}$ emu/mol, has been subtracted from the total susceptibility, and the resulting difference is normalized by the mole fraction of Ce ions. For $x \geq 0.5$, the effective moment $\mu_{eff}$, obtained by a linear fit to data above 200 K, is close to that of the Hund's rule moment of 2.54 $\mu_B$/Ce, and the paramagnetic Curie–Weiss temperature $\theta_P$ changes very little. The negative sign of $\theta_P$ indicates an antiferromagnetic interaction, consistent with an antiferromagnetic ground-state of CeIn$_3$. For $x = 0.2$, the effective moment exceeds the Hund's rule value, which is physically unreasonable. Presently, we do not understand the origin of this larger moment but suspect that it might be due to inaccuracies in subtracting two high temperature values of susceptibility that are comparable—the total susceptibility of $La_{0.8}Ce_{0.2}In_3$ and that of LaIn$_3$. Figure 2b shows the evolution

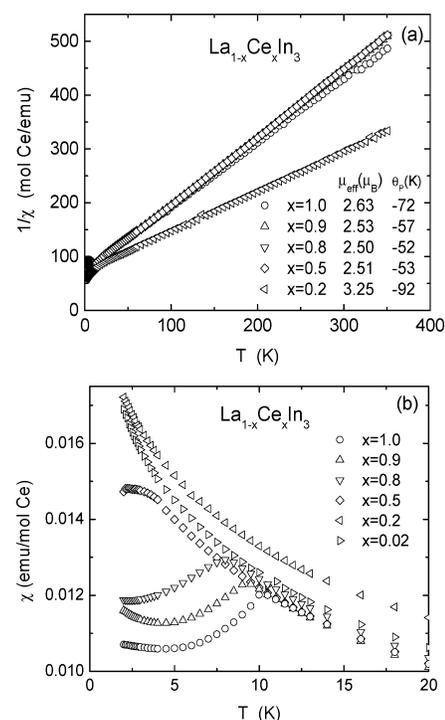

**Figure 2.** (a) Inverse molar magnetic susceptibility as a function of temperature for various values of nominal $x$ in $La_{1-x}Ce_xIn_3$. Values of the high-temperature effective moment $\mu_{eff}$ and paramagnetic Curie–Weiss temperature ($\theta_P$) are given in the caption. (b) Low temperature molar susceptibility for several $La_{1-x}Ce_xIn_3$ compositions. The cusp in susceptibility reflects the onset of antiferromagnetic order.

*Variation of 4f Hybridization Across the La$_{1-x}$Ce$_x$In$_3$ Series*

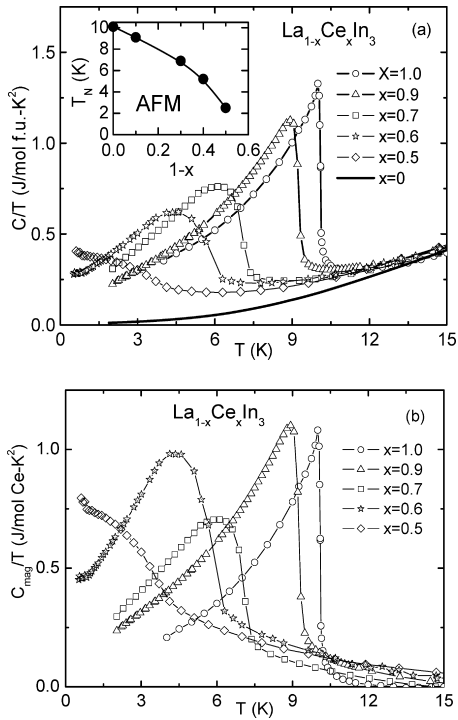

**Figure 3.** (a) Total specific heat, plotted as $C/T$ vs $T$, for La$_{1-x}$Ce$_x$In$_3$ for $0 \leq x \leq 1$. The specific heat is normalized per formula unit mole. The inset shows the suppression of the Néel temperature with decreasing Ce concentration indicating a critical concentration $x_c \approx 0.4$. (b) Magnetic contribution to the specific heat, plotted as $C_{mag}/T$ vs $T$, for Ce$_{1-x}$La$_x$In$_3$ for $0 \leq x \leq 1$. See text for details.

of the low temperature susceptibility and ground-state as a function of $x$. The cusp in $\chi(T)$ near 10 K for $x = 1.0$ signals the development of antiferromagnetic order. With decreasing Ce concentration, the cusp moves to lower temperatures and becomes poorly defined for $x = 0.5$ and is absent above 2 K for $x = 0.2$ and 0.02. This evolution of magnetic order at $T_N$ is confirmed by specific heat measurements and is consistent with a previously report[9] based on susceptibility measurements.

**Specific Heat.** The total specific heat $C$ of La$_{1-x}$Ce$_x$In$_3$, plotted as $C/T$ vs $T$, for $0 \leq x \leq 1$, is displayed in Figure 3a. With decreasing Ce concentration, the anomaly at $T_N = 10$ K in CeIn$_3$ is suppressed, reaching a value of $T_N \approx 2.5$ K at $x = 0.5$ and broadens somewhat, particularly for $x < 0.7$. As shown in the inset of Figure 3a, the Néel temperature decreases monotonically with increasing $x$; these results suggest a critical concentration near $x_c \sim 0.4$. This critical concentration of Ce ions is higher than expected for magnetic order in a simple cubic lattice of Ce ions; for site percolation in a simple cubic lattice, $x_c$ is expected to be 0.31. Magnetic order, however, is mediated by the indirect Ruderman−Kittel−Kasuya−Yosida interaction between Ce ions in the face-centered cubic lattice of La$_{1-x}$Ce$_x$In$_3$; therefore, it is more appropriate to consider the percolation threshold for an fcc lattice that is $x_c = 0.198$.[10] The higher experimental value for $x_c$ suggests that additional effects are present, one of which is hybridization between Ce and In electrons that would favor the more rapid suppression of Néel order that is observed. We note in this regard that the critical concentration of Ce ions required for the first appearance of long-range antiferromagnetism in the layered variant La$_{1-x}$Ce+$_x$RhIn$_5$ is 0.6,[9] which is very close to the theoretically expected value of 0.59 for site percolation on a square lattice[10] and suggests that hybridization of Ce and In electrons does not play such a significant role in this case. The magnetic contribution to the specific heat $C_{mag}$, derived from subtraction of the nonmagnetic LaIn$_3$ contribution from the data, is shown in Figure 3b and is plotted as $C_{mag}/T$ vs $T$, for $0 \leq x \leq 1$. $C_{mag}/T$ is large and increases with decreasing temperature in the paramagnetic state, which is characteristic of heavy-electron materials with strong electron–electron correlations.[11] For each composition, the magnetic entropy at 15 K [$S_{mag} = \int (C_{mag}/T) \, dT$] amounts to $S_{mag}$(15 K) $\sim$ 60–85% of $R\ln 2$, suggesting a crystal-field doublet ground-state of the Ce ion and the presence of hybridization at low temperatures that transfers magnetic entropy from localized to itinerant degrees of freedom. Detailed structural measurements and their modeling, discussed below, suggest that this hybridization is dominated by the Ce f- and In p-orbitals.

**Crystal Structure Determination. Neutron Diffraction.** Because the similar scattering factors for Ce and La may not permit distinguishing these atoms by X-ray powder diffraction (just one electron difference), neutron powder diffraction was carried out on these compounds. Indeed, the elastic neutron cross sections for Ce $(7.718 \cdot 10^{-24}$ cm$^2)$ and La $(5.680 \cdot 10^{-24}$ cm$^2)$ are significantly different to allow us to refine site distributions. Time-of-flight (TOF) neutron diffraction data were collected on the neutron powder diffractometer (NPDF) at the Manuel Lujan Neutron Scattering Center of Los Alamos National Laboratory. This instrument is a high-resolution powder diffractometer located at the end of a flight path 1.32 m from the spallation neutron target. The data were collected at 15 and 295 K using the 148, 119, 90, and 46° banks, which cover a d-spacing range from 0.12 to 7.2 Å. On the basis of the structural arrangement known from the literature for LaIn$_3$ and CeIn$_3$,[12−15] we refined unit cell parameters and isotropic displacement parameters (IDPs) for the various RE compositions and the different temperatures. The structures were refined using the General Structure Analysis System (GSAS), a Rietveld profile analysis program developed by Larson and Von Dreele.[16] Background coefficients, scales factors, isotropic strain terms in the profile function, and sample absorption were also refined for a total of 61 parameters. To elucidate the arrangement and concentration of La and Ce, we allowed the La/Ce occupancies to be refined. At that stage we identified a systematic overestimation of the Ce concentration. To obtain a satisfying Ce/La ratio, it was necessary to release the In occupancy that converged to an unphysical value greater than 1.0. A combined study of observed Fourier and difference Fourier maps gave a hint regarding the possible anomaly. Indeed, elevated residues were present on both sides of the symmetric plane where the In site is located (Wyckoff position *3c*). The anomaly was resolved by the introduction and the refinement of the ADPs for the In site that leads to a slightly large ADP but with an ideal full occupancy on this site. Table 2 summarizes the unit cell parameters and refinement results (displacement parameters) for each composition and temperature. Figure 4 illustrates the observed and calculated neutron diffraction patterns for La$_{0.2}$Ce$_{0.8}$In$_3$ as an example.

**Table 2.** Unit Cell Parameters and Refinement Results for $La_{1-x}Ce_xIn_3$ ($x$ = 0.02, 0.2, 0.5, and 0.8) at 15 and 295 K

| targeted compositions | $La_{0.98}Ce_{0.02}In_3$ | $La_{0.8}Ce_{0.2}In_3$ | $La_{0.5}Ce_{0.5}In_3$ | $La_{0.2}Ce_{0.8}In_3$ |
|---|---|---|---|---|
| | | 15(1) K | | |
| refined cell parameters (Å) | 4.7121(1) | 4.7044(2) | 4.6896(2) | 4.6731(2) |
| refined compositions | $La_{0.98(1)}Ce_{0.02(1)}In_3$ | $La_{0.79(1)}Ce_{0.21(1)}In_3$ | $La_{0.51(2)}Ce_{0.49(2)}In_3$ | $La_{0.18(2)}Ce_{0.82(2)}In_3$ |
| In ADPs (Å$^2$) (100*$U_{11}$/100*$U_{22}$) | 0.52(2)/0.32(1) {1.625} | 0.46(3)/0.25(2) {1.84} | 0.42(4)/0.28(2) {1.50} | 0.50(5)/0.31(3) {1.61} |
| Ce IDPs (Å$^2$) 100*$U_{eq}$ | 0.24(1) | 0.15(1) | 0.13(1) | 0.11(2) |
| $\chi^2$ | 1.840 | 1.937 | 2.274 | 1.751 |
| | | 295(1) K | | |
| refined cell parameters | 4.7312(2) | 4.7237(3) | 4.7130(3) | 4.6912(3) |
| refined Compositions | $La_{0.97(2)}Ce_{0.03(2)}In_3$ | $La_{0.78(2)}Ce_{0.22(2)}In_3$ | $La_{0.51(2)}Ce_{0.49(2)}In_3$ | $La_{0.21(2)}Ce_{0.79(2)}In_3$ |
| In ADPs (Å$^2$) (100*$U_{11}$/100*$U_{22}$) | 2.67(7)/1.13(3) {2.36} | 2.11(8)/0.73(3) {2.89} | 2.21(7)/0.82(6) {2.69} | 0.69(5)/1.39(4) {0.49} |
| Ce IDPs (Å$^2$) 100*$U_{eq}$ | 0.96(2) | 0.95(2) | 0.95(5) | 1.30(8) |
| $\chi^2$ | 1.339 | 1.515 | 3.847 | 1.650 |

*La/Ce in (0, 0, 0) and In in (0, 0.5, 0.5).

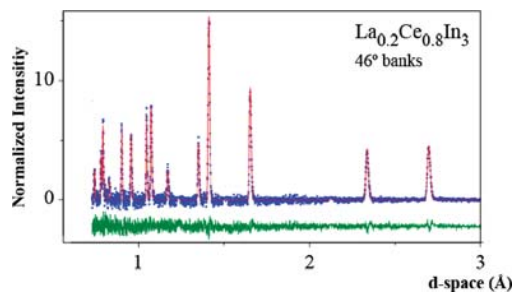

**Figure 4.** TOF neutron powder diffraction patterns for d-spacings between 0.5 and 3 Å for $La_{0.2}Ce_{0.8}In_3$ taken from data obtained with the 46° bank detectors. Observed data are represented by blue crosses; calculated results are represented by the red line. The intensity scale uses arbitrary units. The difference curve is illustrated in green on the same scale.

**X-Ray Diffraction.** Because neutron absorption by In could lead to unrealistic ADPs, X-ray diffraction studies were performed to check the real nature of this phenomena even if the Ce/La ratio could not be refined. X-ray powder diffraction patterns of these phases were recorded using X-rays of energy 98.11 keV ($\lambda$ = 0.126371 Å) at the 6-IDD beam line at the advanced photon source (APS) at Argonne National Laboratory in order to perform both Rietveld refinements and PDF analyses. Data were collected at room temperature using a circular image plate camera (Mar345) 345 mm in diameter. The samples were exposed for 15 s, and to improve the statistic, six runs were taken and combined.

The purpose of these Rietveld refinements was exclusively the investigation of these large ADPs and not the Ce/La ratio previously refined based on the neutron data. Therefore, all refinements have been carried out using Ce as fully occupying the RE site. Final results show a similar anisotropic displacement behavior but with even larger magnitude. Table 3 lists the X-ray Rietveld refinements results, and Figure 5 illustrates the observed and calculated neutron diffraction patterns for $La_{0.2}Ce_{0.8}In_3$ as an example.

**Structural Discussion and PDF Analysis.** The members of the series of $La_{1-x}Ce_xIn_3$ ($x$ = 0.02, 0.2, 0.5, or 0.8) crystallize in the simple cubic $AuCu_3$ structure type. The structural refinements do not show any evidence of ordering between the La and Ce. However, as shown in Figure 6, both neutron and X-ray diffraction analyses have clearly found a strong ADP on the In site. Depending on the La/Ce concentration, two opposite situations are observed. In the case of La-rich compounds, the In could be represented as an ellipsoidal elongation whereas in the Ce-rich case the In corresponds to a flat butterfly-like deformation. The inset in Figure 6 represents a calculated Fourier map which shows that this flat deformation is not circular and could be better described by an anharmonic treatment of the displacement parameters. However, because GSAS does not support such anharmonic analysis, the refinement has been reproduced using the Jana2006 software,[17] which could treat both TOF data and anharmonic treatment of the displacement parameters. To satisfy the symmetry restrictions, a fourth-order Gram−Charlier treatment of the RE and the In atomic displacements was then carried out, a treatment that treats the anisotropic displacement of atoms as anharmonic.[18] An anharmonic displacement has been observed for both sites. However, because of the low amplitude of the RE displacement and the number of parameters added for such description, we have chosen to keep only these extra parameters for the In sites that are significant for the Ce-rich compounds. For example, for $La_{0.2}Ce_{0.8}In_3$ the tensor element $D_{1111}$ is equal to 0.022(1) (Tensors elements $D_{ijkl}$ are multiplied by $10^6$). These values decreased to become negligible as the La concentration increased.

It is interesting to note that the magnitude of these deformations is multiplied by a factor of two or three between X-ray data and neutron data. A stronger observed deformation for the In site in the case of X-ray data is consistent with an electronic origin. Indeed, the X-ray beam is probing the electronic environment surrounding the atoms, whereas the neutron beam gives information on the core position of these atoms. Ideally, we could schematically separate and quantify the electronic contribution of such a deformation if we postulate that the atom core is not following the electronic cloud, which should be true to some extent. Thus, it is still questionable if the effect observed is due to the vibration of the atoms or due to a static disorder.

ADPs determined in the crystal diffraction studies could represent either time-averaged or lattice-averaged probability. The usual way to distinguish between vibration and static disorder is by studying temperature dependence. On the other hand, at a given temperature it is also possible to differentiate between these two scenarios by comparing neutron diffraction data and X-ray diffraction data because it is possible to analyze either the core or the shell of the atom concerned. However, both approaches were successful in discerning unequivocally the origin of ADPs, and it seems possible that both phenomena could occur simultaneously even if one or the other is leading.

To have a better picture of the mechanism responsible for the large In ADPs, we performed PDF analyses to investigate the local structure. A combined study of Rietveld and PDF analysis is generally helpful to provide information to the long, medium, and short-range ordering in a structure. Indeed, Rietveld analysis determines only the average long-range structure because it only

---

*Variation of 4f Hybridization Across the La$_{1-x}$Ce$_x$In$_3$ Series*

**Table 3.** Unit Cell Parameters and Refinement Results for La$_{1-x}$Ce$_x$In$_3$ ($x$ = 0.2, 0.5, and 0.8) at RT from X-ray Rietveld Refinements

| targeted compositions | La$_{0.8}$Ce$_{0.2}$In$_3$ | La$_{0.5}$Ce$_{0.5}$In$_3$ | La$_{0.2}$Ce$_{0.8}$In$_3$ |
|---|---|---|---|
| refined cell parameters (Å) | 4.7044(2) | 4.6896(2) | 4.6731(2) |
| In ADPs (Å$^2$) (100*$U_{11}$/100*$U_{22}$) | 2.52(7)/0.72(7) {3.5} | 3.32(7)/1.98(6) {1.68} | 0.07(1)/0.94(5) {0.07} |
| Ce IDPs (Å$^2$) 100*$U_{eq}$ | 0.99(6) | 1.35(6) | 1.60(6) |
| $\chi^2$ | 1.937 | 2.274 | 1.751 |

* La/Ce in (0, 0, 0) and In in (0, 0.5, 0.5).

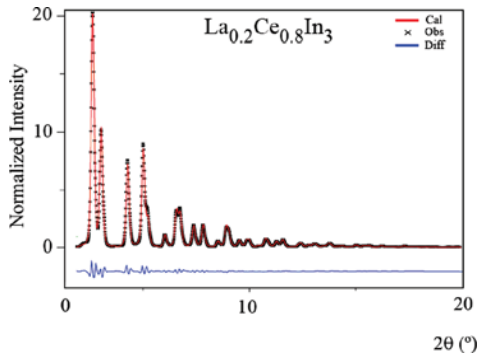

**Figure 5.** X-ray powder diffraction patterns for 2$\theta$ between 0 and 20° for La$_{0.2}$Ce$_{0.8}$In$_3$. Observed data are represented by crosses; calculated results are represented by the red line. The intensity scale uses arbitrary units. The difference curve is illustrated in blue on the same scale.

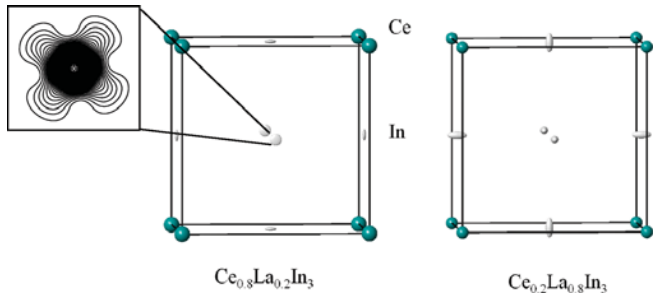

**Figure 6.** Representation of the unit cell for La$_{0.2}$Ce$_{0.8}$In$_3$ and La$_{0.8}$Ce$_{0.2}$In$_3$. The blue and grey atoms represent the RE atoms and In atoms, respectively. The inset shows the calculated Fourier map on the In site based on the X-ray refinement.

takes into account the intensity and the position of the Bragg peaks, whereas PDF analysis also takes into account information contained in the diffuse scattering.

PDFs were corrected for background, incident neutron spectrum, absorption, and multiple scattering and were normalized using the vanadium spectrum to obtain the total scattering structure factor $S(Q)$, using the program PDFgetN.[19] The PDF radial distribution function $G(r)$ was obtained from $S(Q)$ via the following Fourier transform,

$$G(r) = 4\pi r[\rho(r) - \rho_0] = \frac{2}{\pi}\int_0^{Q_{max}} Q[S(Q)-1]\sin Qr\, dQ, \quad (1)$$

where $\rho(r)$ and $\rho_o$ are the atomic pair number density and average number density, respectively. The data were terminated at a value $Q_{max} = 20$ Å$^{-1}$. PDF modeling was carried out using the program PDFGui.[20] The refined results obtained from Rietveld were used as a starting structural model. Two different models were considered in order to elucidate "static displacement" vs "dynamic displacement". The first model is an anisotropic treatment of the In displacement parameter (dynamic effect), whereas the second one is a random relaxation of the In position on both sides of the [CeIn] plane in order to obtain a similar picture, on average. Both models have been refined with the same number of parameters and drastically improve the results obtained by considering an In isotropic displacement parameter. In Figure 7 are representative PDF fits for both models up to $r = 20$ Å. Notice that the PDF analyses indicate the shift of the In atoms is less than a tenth of an Angstrom. The quality of both fits is extremely good and, based on the statistical $R$ value, the dynamic effect vs static effect seems to be marginally preferred. This conclusion is consistent with the hypothesis that electronic behavior leads, at least partially, to these atomic deformations.

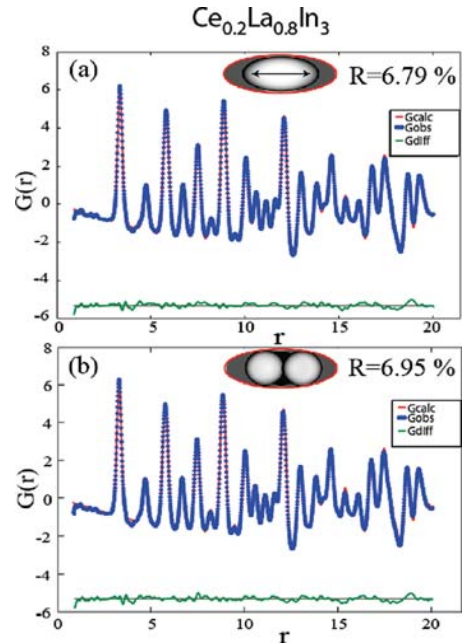

**Figure 7.** PDF refinements over the $r$ range from 1 to 20 Å for La$_{0.8}$Ce$_{0.2}$In$_3$ (a) using a dynamic displacement model and (b) a static displacement model (see text for details). The difference curves between observed and calculated are shown below each graph on the same intensity scale.

**Electronic Structure Calculations.** Band structure calculations have been performed to explore possible electronic reasons for such anisotropic elongation of the In site. Electronic structures were calculated using the self-consistent tight-binding LMTO method (Linear Muffin Tin Orbitals) with the atomic spheres approximation (ASA),[21] using the LMTO version 4.7 program.[22] In the LMTO approach, the density functional theory is employed utilizing the local density approximation (LDA) for the exchange correlation energy.[23] All relativistic effects except spin–orbit coupling were taken into account using a scalar relativistic approximation.[24]

The basis set in these calculations included the 6s, 5d, and 4f orbitals for La and Ce and the 5s and 5p for In. The **k**-space integrations to determine self-consistent charge densities, densities of states (DOS), and crystal orbital Hamilton populations (COHP)[25] were performed by an improved tetrahedron method[26] on a grid of 84 irreducible **k**-points. The Fermi level was selected as the energy reference.

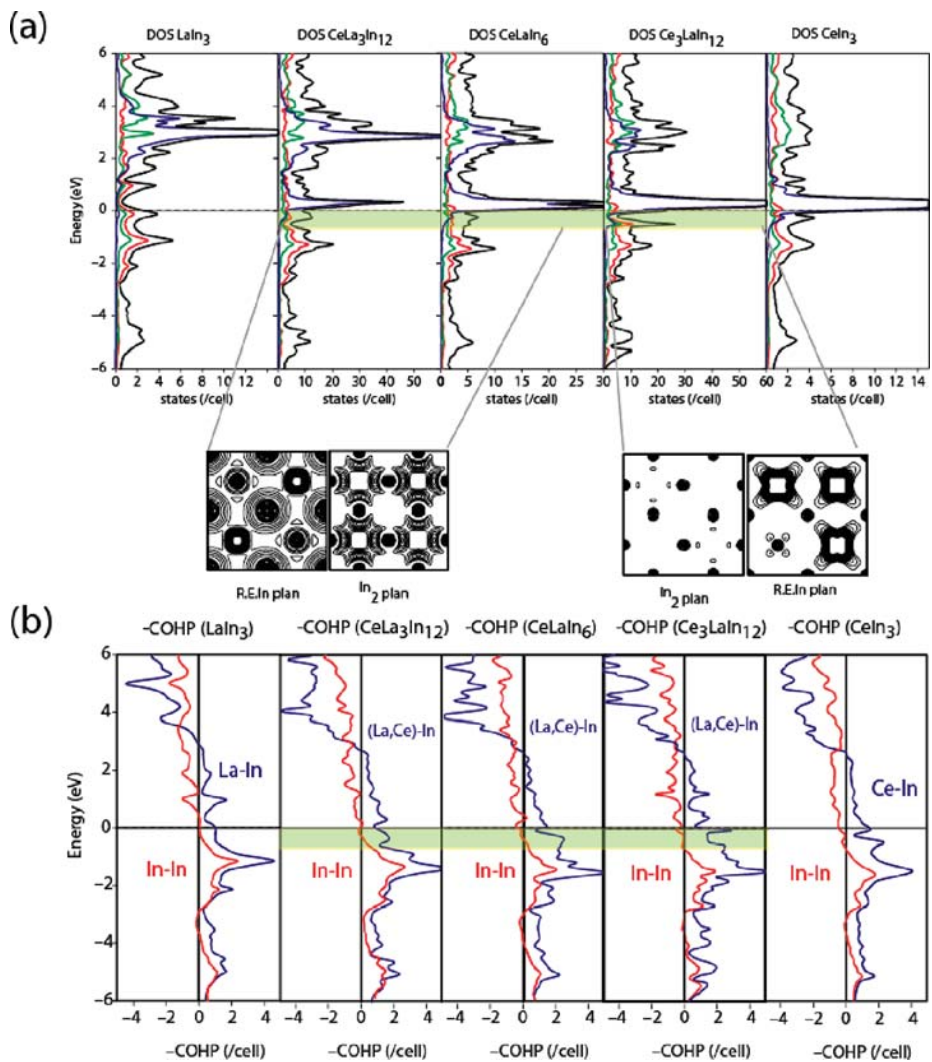

**Figure 8.** (a) TDOS (black curve) and different PDOS for $LaIn_3$, $CeLa_3In_{12}$, $CeLaIn_3$, $Ce_3LaIn_{12}$, and $CeIn_3$. (Ce,La)-f PDOS, (Ce,La)-f PDOS, and In-p PDOS are represented in blue, green, and red, respectively. Below, four electron density maps for the energy windows highlighted in green (RE In plan and $In_2$ plan for the Ce-rich cases and La-rich cases). (b) In-In COHP (red curve) and (Ce,La)-In COHP (blue curve) for $LaIn_3$, $CeLa_3In_{12}$, $CeLaIn_3$, $Ce_3LaIn_{12}$, and $CeIn_3$.

To perform band structure calculations on different members of the $La_{1-x}Ce_xIn_3$ series, different models have been studied that are, for some of them, approximations to the actual atomic arrangement because we were not able to fully treat the RE site as a Ce/La mixed site. Indeed, different primitive settings have been used to create an "artificial" Ce/La ordering. Note that no additional superstructure reflections have been observed, which means that the Ce and La are not ordered. Nevertheless, such orderings in the cases of $La_{0.8}Ce_{0.2}In_3$, $La_{0.5}Ce_{0.5}In_3$, and $La_{0.2}Ce_{0.8}In_3$ correspond to close approximations to the "real" structures. Calculations, therefore, were made for $CeLa_3In_{12}$, $CeLaIn_6$, and $Ce_3LaIn_{12}$, respectively, as approximations to the actual compositions.

Figure 8a illustrates the total DOS (TDOS) and some partial DOS (PDOS) for $LaIn_3$, $La_{0.8}Ce_{0.2}In_3$ ($CeLa_3In_{12}$), $La_{0.5}Ce_{0.5}In_3$ ($CeLaIn_6$), $La_{0.2}Ce_{0.8}In_3$ ($Ce_3LaIn_{12}$), and $CeIn_3$ 6 eV below and above the Fermi level. (Ce, La)-f PDOS, (Ce,La)-d PDOS, and In-p PDOS are represented in blue, green, and red, respectively. From the TDOS and PDOSs we can easily assign the nature of the electronic contributions. In the valence band, two major broad peaks centered at −5 eV and −1 eV can be recognized. Whereas the first one is mostly due to In-s states, the second one just below the Fermi level is due to In-p states as well as RE-d and RE-f orbitals, which imply that electronic interactions may occur between these orbitals. In particular, we notice possible hybridization between conduction electrons and f-electrons which is typical physical behavior for heavy-fermion materials. In the conduction band of Ce-containing alloys, we recognize two peaks centered at 3 eV and just above the Fermi level that correspond to the RE-d and RE-f unoccupied states.

In order to elucidate the nature of the bonding in these materials in relation to their large ADPs, COHP analysis (Figure 8b) and electronic charge density maps in an energy window just below the Fermi level have been performed (green highlighted region in Figure 8a). Indeed, these last occupied states should reflect the nature of the bondings between atoms in the structure. For the La-rich compounds up to $LaCeIn_6$, the maps show that these occupied states correspond to some weak RE-(d,f)/In-p bonding interactions and mainly In-p/In-p bonding interactions. The second map nicely shows the shape of In-p orbitals and how they point toward the center of the unit cell in order to maximize their overlap and, therefore, the bonding character. This representation is in agreement with the observed In-ADP. Indeed, the elongation is also pointing toward the center of the unit cell and so in the same direction as the In-($p_x$, $p_y$) orbitals shown in the electronic charge density map. For the Ce-rich compounds, the situation is different because the calculations show a relative nonbonding character between the In atoms. However, we observe stronger bonding interactions between the In-p orbitals and the RE-(d,f) orbitals. Such hybridizations could be related to the anharmonic behavior observed by diffraction of the In atoms with the density pointing toward the RE atoms. Notice that for Ce$_3$LaIn$_{12}$ and, especially, CeIn$_3$ the deformation is mostly due to the Ce-f orbitals, as compared to the Ce-d orbitals.

---

(25) Dronskowski, R.; Blöchl, P. E. *J. Phys. Chem.*, **1993**, *97*, 8617.
(26) Blöchl, P. E.; Jepsen, O.; Andersen, O. K. *Phys. Rev. B*, **1994**, *B49*, 16223.

COHP analyses presented in Figure 8b revealed the same trend as observed with the charge density maps. Indeed, as the Ce/La ratio increases we observe that the In−In bonding interactions just below the Fermi level decrease, and simultaneously, the In-RE bonding interactions increases. Such COHP studies have already been useful for understanding In displacement in relation to a second-order Jahn–Teller instability[27] or to elucidate stability in energy associated with extended bonding subnetwork[28,29]

## SUMMARY

Structural characterization of the series La$_{1-x}$Ce$_x$In$_3$ ($x =$ 0.02, 0.2, 0.5, or 0.8) shows a specific atomic elongation on the In site depending of the nature of the RE substitution. Indeed, for the La-rich compounds, the In ADP presents an ellipsoidal elongation, whereas, for the Ce-rich compounds, a butterfly-like distortion is observed. Rietveld and PDF analyses using both X-ray and neutron diffraction tend to indicate an electronic origin of such deformation. Electronic structure calculations identify influences of hybridization between the Ce-f orbitals and the In-p orbitals as a possible reason.

**Acknowledgment.** This work has benefited from the use of NPDF at the Lujan Center at Los Alamos Neutron Science Center, funded by the U.S. Department of Energy (DOE) Office of Basic Energy Sciences. Los Alamos National Laboratory is operated by Los Alamos National Security LLC under DOE Contract No. DE-AC52-06NA25396. The upgrade of NPDF has been funded by NSF through grant No. DMR 00-76488. The authors are also grateful to Alfred Kracher of the Ames Laboratory for performing electron microprobe analyses on the samples. Ames Laboratory is operated for the U.S. Department of Energy by Iowa State University under Contract No. W-7405-ENG-82 and is supported by the Materials Sciences Division of the Office of Basic Energy Sciences of the U.S. Department of Energy.

**Supporting Information Available:** Neutron crystallographic files in CIF format for the structure determination of La$_{1-x}$Ce$_x$In$_3$ ($x =$ 0.02, 0.2, 0.5, or 0.8). This material is available free of charge via the Internet at http://pubs.acs.org.